\documentclass{article}
\usepackage{graphicx} 
\usepackage{subcaption}
\usepackage{url} 
\usepackage{xargs} 
\usepackage{comment}
\usepackage[linesnumbered,ruled,vlined]{algorithm2e}
\newtheorem{example}{Example}

\usepackage[colorinlistoftodos,prependcaption,textsize=tiny]{todonotes}
\usepackage{amssymb}
\usepackage{amsmath}
\usepackage{authblk}

\newcommandx{\paul}[2][1=]{\todo[linecolor=red,backgroundcolor=red!25,bordercolor=red,#1]{Paul: #2}}

\SetCommentSty{mycommfont}
\SetKwProg{Fn}{Function}{:}{} 

\title{A Fast Ethereum-Compatible Forkless Database}

\date{\vspace{-5ex}}

\begin{document}


\author[]{Herbert Jordan}
\author[]{Kamil Jezek}
\author[]{Pavle Suboti\'{c}}
\author[]{Bernhard Scholz}
\affil[]{Sonic Labs}

\maketitle
\begin{abstract}
The \emph{State Database} of a blockchain stores account data and enables authentication.
Modern blockchains use fast consensus protocols to avoid forking, improving throughput and finality. 
However, Ethereum’s StateDB was designed for a forking chain that maintains multiple state versions.
While newer blockchains adopt Ethereum’s standard for DApp compatibility, they do not require multiple state versions, making legacy Ethereum databases inefficient for fast, non-forking blockchains. 
Moreover, existing StateDB implementations have been built on key-value stores (e.g., LevelDB), which make them less efficient.

This paper introduces a novel state database that is a native database implementation and maintains Ethereum compatibility while being specialized for non-forking blockchains. Our database delivers ten times speedups and 99\% space reductions for validators, and a threefold decrease in storage requirements for archive nodes.
\end{abstract}

\section{Introduction}

Blockchain databases have become a focal point for achieving high throughput and low finality in the rapidly evolving landscape of blockchain technology.
The introduction of sophisticated consensus protocols~\cite{BairdL20,tendermint,lachesis}, has fundamentally transformed modern blockchains.
These advancements have enabled a shift from the traditional forking model (where a block may have multiple successors that are eventually pruned except for one) to efficient forkless chains. 
A forkless chain maintains a single linear block history, effectively eliminating the complications associated with resolving competing branches through rules such as the longest-chain rule~\cite{Nakamoto2009BitcoinA}.

However, for achieving real-world traction, these modern blockchains must remain compatible with legacy ecosystems such as Ethereum, which remains one of the most widely used platforms in the blockchain space. 
Hence, non-forking blockchains continue to rely on Ethereum’s state database for storage and authentication, even though its design is ill-suited for forkless chains.
Consequently, the storage layer has emerged as the primary performance bottleneck, with recent studies showing that it accounts for up to 75\% of block processing time~\cite{abs-1910-11143}.
Addressing these storage inefficiencies is essential to unlocking the full potential of modern blockchain solutions and enhancing scalability, speed, and the overall user experience.

The Ethereum blockchain database uses Merkle-Patricia Tries (MPT) to handle authentication and multiple versions of its state~\cite{mpt, kamil}. 
In this structure, each leaf node contains a value, such as an account balance, while the path from the root to the leaf encodes the corresponding key, like an account address. 
This encoding is done by splitting the key into segments, with each inner node of the trie representing a portion of the key.


To support versioning, the MPT functions as an \emph{append-only} data structure.
When a leaf node's value is updated, a new leaf is created to hold the updated value, and all nodes along the path from the root to the original leaf are cloned. 
This design allows multiple root nodes to coexist, each representing a distinct state version.
Consequently, the Merkle-Patricia Trie naturally supports forks and maintains various versions of the state for blocks that may later emerge or be discarded.

Beyond storing the state, the state database also provides \emph{authentication}. 
Each innner node stores the cryptographic hash of the concatenated contents of its children, allowing unchanged sub-trees to be shared across different states. 
The root hash serves as a cryptographic commitment to a specific version of state.
Ethereum's hash compatibility is important in the context of witness proofs.
Witness proofs are compact cryptographic objects that allow a verifier to check the correctness of a sub-state even outside of the chain.
The substate can be a specific piece of state such as an account balance or a contract storage value without requiring access to the full blockchain~\cite{mpt}. 
The witness proof consists of the values being verified together with their paths of hashes that links them to the root of the version in which they reside in.
Using only the root hash and the witness proof, any party can independently recompute and verify the correctness of the sub-state. 
Witness proofs are fundamental building block for modern blockchains, including light clients~\cite{chatzigiannis2022sok},  oracles \cite{al2020trustworthy}, stateless blockchain designs~\cite{10.1007/978-3-031-47751-5_4},  rollups~\cite{thibault2022blockchain} and cross-chain bridges~\cite{11080119}. 
Maintaining Ethereum compatibility for witness proofs is crucial for ensuring cross-chain operability. 



For forkless blockchains, the ability to model non-linear multi-versioning is unnecessary. 
Maintaining multiple versions of a block state leads to multiple state roots, which is costly because it requires frequent pruning (i.e., the disposal of obsolete versions/states) as well as data duplication. 
Another weakness of current implementation is a system aspect.
Legacy implementation of Ethereum compatible state databases (e.g. the Ehereum Foundation's client geth\cite{go-ethereum}) use key-value stores implemented as an ordered data structure 
(such as B-trees and LSM trees), which exhibit a logarithmic worst-case runtime complexity for accessing data.
Building a state database  on top of a key value store (e.g. LevelDB\cite{leveldb}) results in a \emph{read amplification}~\cite{readamp} problem.
The keys in such implementation represent the hash of a node in the MPT, and the value represent the RLP encoded content of a node in the MPT trie.
Such organization becomes inefficient due to the size of the hashes which is a 256-bit value. 



In this paper, we introduce a database design for forkless blockchains that provides significant performance improvements while also retaining compatibility with Ethereum authentication and its witness proofs.
In that regard, our database enables modern forkless blockchains to increase throughput and reduce storage costs while remaining fully Ethereum-compatible for decentralized applications (DApps) with respect to root hashes and witness proofs.

We present two versions of our database, namely, \textsc{LiveDB} and \textsc{ArchiveDB}, each specialized for  different roles in a blockchain.
The \textsc{LiveDB} is a database for validator and observer roles. 
To avoid unnecessary copying in existing databases, the \textsc{LiveDB} uses a mutable data structure that allows data to be overwritten and destroyed, enabling it to evolve to the next state without retaining the old one. 
Thus, pruning is performed intrinsically at no additional cost within \textsc{LiveDB}. 


To ensure authentication compatibility, \textsc{LiveDB} offers the same hash and witness proofs as the Ethereum blockchain. 
This is achieved by providing an interface that allows the storage and retrieval of key-value pairs in a way that aligns with Ethereum's design. 
Additionally, \textsc{LiveDB} maintains a data representation in memory that mirrors the structure of the Ethereum database and utilizes the same RLP encoding, hashing functions, and other mechanisms.



The \textsc{ArchiveDB} is designed specifically for archive node roles. 
It utilizes an append-only, continuously growing data structure that effectively compresses a forkless chain of blocks.
This approach minimizes the need for repeated key copying and hash recalculations, which can occur when values change at the leaf nodes of a database.
The \textsc{ArchiveDB} allows only one subsequent version for each block, enabling it to store changes between two consecutive blocks more efficiently using a delta update style. Similar to the \textsc{LiveDB}, the \textsc{ArchiveDB} ensures full hash compatibility with the Ethereum blockchain.

We have implemented \textsc{LiveDB} and \textsc{ArchiveDB} in the Sonic Blockchain and evaluated their utility on real-world transactions in a public testnet.
Our evaluation shows that \textsc{LiveDB} reduces the storage footprint on the Sonic blockchain by 99\% while increasing throughput almost tenfold compared to using Ethereum's Geth client~\cite{go-ethereum}.

Our contributions are as follows:
\begin{itemize}
\item We provide two forkless databases, \textsc{LiveDB} and \textsc{ArchiveDB}, that are specialized for validators/observers and archive nodes, respectively.
\item We enhance runtime while reducing memory and disk usage by replacing hash-based addressing of trie nodes to linear-number addressing.
\item We overcome the read amplification problem by using a file-mapped array instead of using Key-Value stores (e.g. LevelDB).
\item We reduce space by employing more efficient hash encodings.
\item We solve the pruning problem by employing destructive updates in the \textsc{LiveDB}.
\end{itemize}


This paper is structured as follows.
In Section~\ref{sec:livedb} we describe the \textsc{LiveDB} configuration.
In Section~\ref{sec:archivedb} we describe the ArchiveDB configuration.
In Section~\ref{sec:opt} we describe further optimizations in the implementation.
In Section~\ref{sec:eval} we evaluate the \textsc{LiveDB} and \textsc{ArchiveDB}.
In Section~\ref{sec:related} we contrast our contribution to related work.
We conclude in Section~\ref{sec:conclusion}.

\section{LiveDB}
\label{sec:livedb}
In this section, we describe the \texttt{LiveDB} database and discuss its design decisions. 

\subsection{Overview}

Blockchains encode state transitions through transactions recorded in each block.
These transitions typically involve updates to account balances, deployment and execution of smart contracts, and modifications to contract-specific program variables.
Although the transaction history provides a complete and immutable record of all state changes, reprocessing the entire chain to reconstruct the current state would be computationally prohibitive.
To address this, blockchain systems maintain a dedicated state database that reflects the latest state resulting from the execution of all prior transactions.
This database serves as an efficient lookup structure, enabling rapid access to the current or previous state required for validating and executing new transactions.
This component is performance critical as it determines how fast the state can progress and thus has a direct impact on performance in terms of the number of processed transactions per second. 

At the interface/semantic level, the state database in blockchain systems functions as a key-value store, where keys correspond either to 20-byte account addresses or with 32-byte smart contract storage slots.
The associated values represent account metadata—such as balances and nonces or the contents of individual storage slots.
During transaction execution, this interface is used to retrieve and update state elements relevant to accounts and contracts.
The state database serves two principal roles: first, it provides access to the most recent state required to validate and apply new transactions at the tip of the chain; second, it enables reconstruction of the state as it existed at any specific block height, thereby supporting historical queries and analysis of prior chain states.

In most blockchain implementations, both real-time state access and historical state reconstruction are served by a unified, general-purpose database.
While this approach simplifies system architecture, it introduces limitations due to the divergent performance and access requirements of these two roles.
The recent state must be available with minimal latency to support transaction execution and block validation at the chain’s tip.
In contrast, historical queries—used for auditing, analytics, or archival inspection—do not demand low-latency access but often benefit from asynchronous retrieval mechanisms.
Moreover, the historical state may be entirely irrelevant in certain operational contexts, such as validator nodes focused solely on forward progression.
This mismatch suggests that separating state and history into distinct subsystems may yield architectural and performance advantages.
As a solution we propose two separate databases the \textsc{LiveDB} and the \textsc{ArchiveDB}

\begin{figure}
    \centering
    \includegraphics[width=0.4\linewidth]{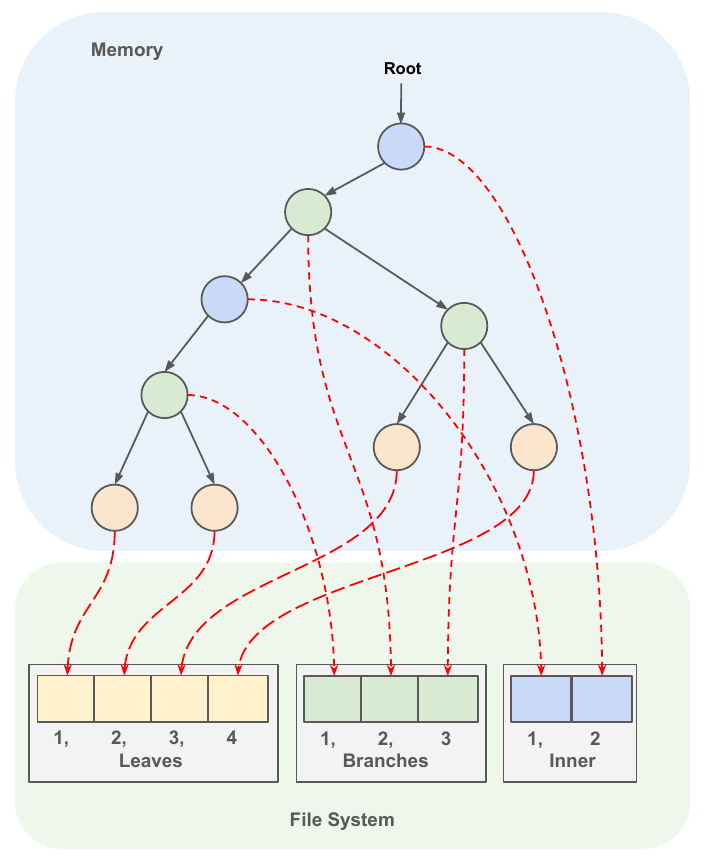}
    \caption{\textsc{LiveDB} Architecture}
    \label{fig:overview}
\end{figure}

The \textsc{LiveDB} spans two components: an in-memory overlay and the disk storage persistence layer.
We depict the high level architecture in Figure~\ref{fig:overview}.
The memory overlay of \textsc{LiveDB} is a tree structure, while the persistence layer flattens the nodes in a file mapped array for persistence. 
The memory overlay contains only a fraction of the entire state that is actually needed for the current context: typically, currently being processed transactions, blocks, as well as RPC queries for witness proofs and other information about accounts.
In the proceeding subsection we describe each of these layers.

\subsection{In-Memory Representation}
The in-memory overlay provides an external interface for reading and writing data.
We denote this layer as an abstract data type we refer to as the \textsc{LiveDB}:
$\textsc{LiveDB} = \langle \mathcal{T}, \texttt{Get}, \texttt{Put}, \texttt{Delete}, \texttt{Commit} \rangle$ consisting of a rooted tree $\mathcal{T}$ and operations  \texttt{Get}, \texttt{Put}, \texttt{Delete} and \texttt{Commit}.

We define $\mathcal{T} = \langle r, N, E \rangle$ where $r \in N$ is the unique root node, $N$ is the set of nodes and $E$ is the edge relation $N \times N \rightarrow r$ between nodes such that  $\mathcal{T}$ forms a rooted tree.
A path of the tree is a set of ordered nodes $P \subseteq N$ such that $P$ forms a \emph{directed path} w.r.t the edge relation $E$.
We further differentiate $N$ into three types: leaf, branch and inner denoted by $L, B, I$, respectively, such that  $L \cup B \cup I = N$.
Nodess have fields that can be accessed using the notation $node.field$. 
We define the fields for each node type as follows. 
For leafs we let $l = \langle key, value, dirty, id\rangle \in L$ where $key \in u256$ is an up-to 32 byte key suffix, $value \in V$ is the stored value and $dirty \in \mathbb{B}$ is a flag that determines if  the value has been updated, $id \in \mathbb{N}$ is an id.
For branch nodes, $b=\langle C, H, dirty, id \rangle \in B$; $C$ is the list of children ids, where the index $\in u16$ in this list is a half-byte partial key, $H$ is the list of hash of children and dirty is the $dirty \in \mathbb{B}$ flag signifying the hash fields have been changed, $id \in \mathbb{N}$ is an id.
For $i = \langle key, c, h, dirty, id \rangle \in I$ where $key \in u256$ is an up-to 32 byte shared path of the key,  $c$ is a single child, $h$ is the child hash and $dirty \in \mathbb{B}$ so the dirty flag, $id \in \mathbb{N}$ is an id.
The $id$ field is auto generated on node construction by the \texttt{GetID} operator in the persistence layer and thus is implicit.
The key in every node enables the addressing of values and resembles a trie data structure as shown in the following example. 

\begin{figure}
    \centering
    \includegraphics[width=0.4\linewidth]{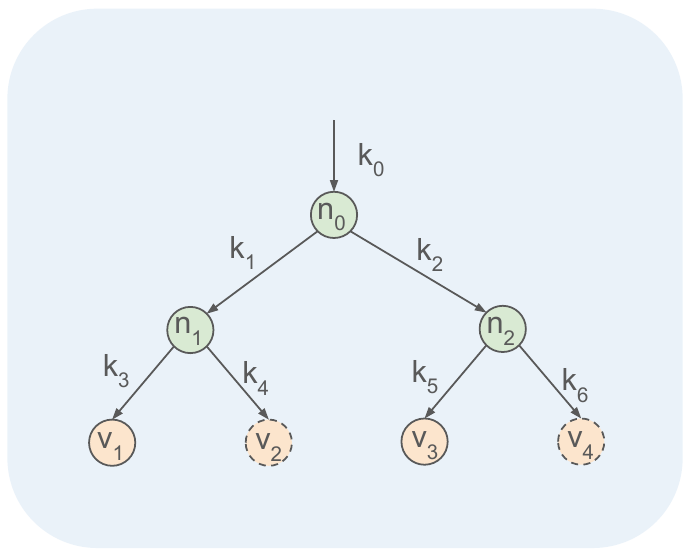}
    \caption{LiveDB with keys and C fields}
    \label{fig:livedb}
\end{figure}

\begin{example}
    Consider the \textsc{LiveDB} in 
    Figure~\ref{fig:livedb} where keys and edges are depicted. For instance, $n_0 = r$ and $n_1 \in B$ and $v_2 \in L$. $n_0.C = n_1::n_2$ and $n_1.C = v_1::v_2$. 
    Given a path consists of these 3 nodes $n_0$, $n_1$, and the $v_2$ lets assume such that $n_0.key = k_0, n_1.key = k_1, v_2.key=k_4 $. Then by concatenating ($\circ$) the keys of each node we form the full key. Thus in our example, the full key $k_0 \circ k_1 \circ k_4 = k_0k_1k_4$ maps to $v_2.v$ i.e., $k_0k_1k_4 \mapsto v_2.v$.
\end{example}

The operators  $\texttt{Get}: \mathbb{N} \rightarrow V$, and $\texttt{Put}: \mathbb{N} \times V \rightarrow N$ and 
$\texttt{Delete}:\mathbb{N} \rightarrow N$ and $\texttt{Commit}: () \rightarrow u256$ are defined in 
Algorithms~\ref{alg:get}, \ref{alg:put}, \ref{alg:del} and \ref{alg:hashcomp}, respectively. 


The \texttt{Get} operation (Algorithm~\ref{alg:get}) operates recursively from the root of the trie, effectively navigating through it based on the type of each encountered node. 
Upon loading a node from storage, the algorithm distinguishes whether it is a leaf, branch, or inner node, determining the appropriate descent strategy.
For inner nodes, it accurately verifies the shared prefix between the input key and the node’s stored path, consuming this prefix before continuing the traversal. 
Branch nodes, on the other hand, consume a single nibble of the key to select the corresponding child node.
When the algorithm reaches a leaf node, it checks whether the remaining portion of the key matches the leaf’s stored key. If there is a match, the associated value is returned. However, if the key fails to match the expected path segment at any point, or if the traversal leads to a missing child node, the lookup process is halted, and the value is deemed empty.


\SetKwFunction{FGet}{\texttt{Get}}
\SetKwFunction{FPut}{\texttt{Put}}

\SetKwFunction{FGetAux}{\texttt{Get\_aux}}
\SetKwFunction{FPutAux}{\texttt{Put\_aux}}

\begin{algorithm}
\caption{\texttt{Get} Operator Computation\label{alg:get}}
\KwData{$\mathcal{T}$: Rooted tree}

\Fn{\FGet{key}}{
 \Return \FGetAux($key$, $\mathcal{T}.r$)\;
}

\Fn{\FGetAux{$key, nodeId$}}{

\Begin{
  \tcp{Call to persistence layer} 
  $node \gets \texttt{Load}(nodeId)$\;
  \Switch{$node$}{
    \Case{$node \in L$}{
       \If{$node.key = key$}{
            \Return $node.value$\;
        }
        \Return $nil$\;
    }
    \Case{$node \in B$}{
        \If{$node.C[key[0]] = nil$}{
            \Return $nil$\;
        }
        \Return \FGetAux{$node.C[key[0]]$, $key[1:]$}\;
    }    
    \Other{
        \For{$i \gets 0$ \KwTo $len(node.key) - 1$}{
            \If{$node.key[i] \neq key[i]$}{
                \Return $nil$\;
            }
        }

        \Return \FGetAux{$node.c$, $key[len(node.key):]$}\;    
    }
  }
}

}
\end{algorithm}


The \texttt{Put} operation (Algorithm~\ref{alg:put}) introduces structural complexity because it may need to modify the trie when the path corresponding to the input key is not fully present. 
As with the \texttt{Get} procedure, the algorithm traverses the trie by interpreting the key according to the semantics of each node type.
In the simplest case, the traversal reaches a leaf node whose key matches the input, and the associated value is updated directly. 
However, if the traversal terminates at a node whose key diverges from the input, the algorithm must construct a new subtree to accommodate both the existing and the new key. 
This process involves splitting the conflicting node and introducing a new branch node to resolve the divergence.
Furthermore, if the two keys share a common prefix, an inner node is inserted to represent the shared path segment above the newly created branch.

\enlargethispage{3\baselineskip}
\begin{algorithm}
\caption{\texttt{Put} Operator Computation\label{alg:put}}
\KwData{$\mathcal{T}$: Rooted tree}

\Fn{\FPut{key, value}}{
 \Return \FPutAux($key$, $value$, $\mathcal{T}.r$)\;
}

\Fn{\FPutAux{key, value, nodeId}}{

\Begin{
  \tcp{Call to persistence layer} 
  $node \gets \texttt{Load}(nodeId)$\;
  \Switch{$node$}{
    \Case{$node \in L$}{
      \If{node.key = key}{
         $node.value \gets value$ \tcp*{Mutable update}    
         $node.dirty \gets true$\;
        \Return{node}
      }
      $lcp \gets$ longest common prefix of $node.key$ and $key$\;

      \tcp{This node will split to branch and two children}    
      $branch \leftarrow \texttt{Split}(node)$\;

      \If{$lcp > 0$}{
        \tcp{Prepend with inner node consuming common path}    
        \Return I(key: $key[0:lcp]$, c: $branch$, h: $nil$, dirty: $true$)\;
     }

     \Return $branch$\;
    }
    
    \Case{$node \in B$}{
      $node.C[key[0]] \leftarrow $\FPutAux($node.C[key[0]], key[1:], value$)\;
      \Return $n$\;
    }
    \Other{
      $lcp \gets$ longest common prefix of $node.Key$ and $key$\;

    \If{$pLn = |node.Key|$}{
        $node.c \leftarrow $\FPutAux($node.c, key[lcp:], value)$\;
        \Return $node$\;
    }

   \tcp{This node will split to branch and two children}    
    $branch \leftarrow \texttt{Split}(node)$\;

    \If{$lcp > 0$}{
        \tcp{Prepend with inner node consuming common path}    
        \Return I(key: $key[0:lcp]$, c: $branch$, h: $nil$, dirty: $true$)\;
    }
    \Return $branch$\;
    }
  }
}
}

\end{algorithm}


The \texttt{Delete} operation (Algorithm~\ref{alg:del}) follows the same traversal logic as the \texttt{Put} operation, navigating through the trie while respecting the semantics associated with each type of node.
The complexity of this operation does not arise from finding the key itself, but rather from the necessary structural adjustments after a successful deletion.
Specifically, if the removal of a subtree results in a branch node having only a single child—and that remaining child is either a leaf or an inner node—the branch will be merged with its last child and then removed.
Similarly, when an inner node has a child that is either a leaf or another inner node, the two nodes will be merged to eliminate unnecessary indirection.

\SetKwFunction{FDelete}{\texttt{Delete}}
\SetKwFunction{FDeleteAux}{\texttt{Delete\_aux}}

\begin{algorithm}
\caption{\texttt{Delete} Operator Computation\label{alg:del}}
\KwData{$\mathcal{T}$: Rooted tree}

\Fn{\FDelete{key}}{
 \Return \FDeleteAux($key$, $\mathcal{T}.r$)\;
}
\Fn{\FDeleteAux{key, nodeId}}{
  $node \gets \texttt{Load}(nodeId)$\;
  \Switch{$node$}{
    \Case{$node \in L$}{
      \If{$node.key = key$}{
        \tcp{Call to persistence layer} 
        $\texttt{Delete}(nodeId)$\;
        \Return $nil$
      }
      \Return $node$
    }

    \Case{$node \in B$}{
      $node.C[key[0]] \gets$ \FDeleteAux{$key[1:],\ node.C[key[0]]$}\;

      $nnChildren \gets \text{non-nil children of } node.C$\;
      \If{$ |nnChildren| = 1 \wedge nnChildren[0] \in L \cup I$ }{
           $\texttt{Delete}(nodeId)$\;
           \Return $\texttt{Merge}(node, nnChildren[0])$\;
      }

      $node.dirty \gets true$\;
      \Return $node$\;
    }

    \Other{
      $lcp \gets$ longest common prefix of $node.Key$ and $key$\;

      \If{$lcp < len(node.key)$}{
        \Return $node$
      }

      $node.c \gets$ \FDeleteAux{$key[lcp:],\ node.c$}\;

      \If{$node.c = nil$}{
        \tcp{Call to persistence layer} 
        $\texttt{Delete}(nodeId)$\;
        \Return $nil$
      }

      \If{$node.c \in I \cup L$}{
        $\texttt{Delete}(nodeId)$\;
        \Return $\texttt{Merge}(node, node.c)$\;
      }

      $node.dirty \gets true$\;
      \Return $node$\;
    }
  }
}
\end{algorithm}


After each block is processed, the \texttt{Commit} operation is executed.
The \texttt{Commit} operation invokes the internal procedure \texttt{UpdateHash}, which is responsible for updating the node hashes. 
The \texttt{UpdateHash} function is specified in Algorithm~\ref{alg:hashcomp}. 
It updates all affected node hashes and, consequently, the root hash of the \textsc{LiveDB}.
The algorithm begins at the node whose hash is to be computed, which is typically the root. 
All immediate children of this node are examined, and any dirty nodes are pushed onto a stack. 
The algorithm then proceeds in a depth-first manner by repeatedly popping the top element from the stack and applying the same procedure.
During backtracking, the hash of the current node is computed. 
At this point, the entire subtree rooted at the node has already been processed, since hashes are calculated only after all descendant nodes have been visited. 
This strategy ensures that hashes are computed in a bottom-up manner, layer by layer, thereby avoiding redundant recomputation of inner node hashes.

\SetKwFunction{FCommit}{\texttt{Commit}}
\SetKwFunction{FHash}{\texttt{UpdateHash}}
\SetKwFunction{FPer}{\texttt{Persist}}
\begin{algorithm}
\caption{\texttt{Commit} Operator Computation\label{alg:hashcomp}}
\KwData{$stack$: Stack}
\KwData{$queue$: Queue}
\KwData{$hashes: N \rightarrow u256$}
\KwData{$\mathcal{T}$: Rooted tree}
 \Fn{\FCommit{}}{
 \texttt{UpdateHash}($\mathcal{T}.r$)\;
 \texttt{Persist}($\mathcal{T}.r$)\;
 }
 \Fn{\FHash{r}}{
   queue.push(r)\;
   \tcc{Collect dirty subtree in stack}
   \While{!queue.empty()}{
    $t \gets queue.enque()$\;  
    \For{$c \in t.C$}{
        \If{$c.dirty$}{
           stack.push(c)\;
           queue.push(c)\;
         }
     }
   }

   \tcc{Compute hash from stack}
   \While{!stack.empty()}{
     $n \gets stack.pop()$\;
     \If{$n \in L$}{
       $hashes[n] \gets hash(n)$\;
     }
     \ElseIf{$n \in I \cup B$}{       
       \For{$i \in range(0, |n.C|)$} {
          $n.H[i] \gets hashes[n.C[i]]$\;
       }
       $hashes[n] \gets hash(n)$\;
     }
   }
   \Return{hashes[r]}
}

\Fn{\FPer{n}}{
\Switch{$n$}{
    \Case{$n \in L$}{
      \texttt{Store}(n.id, n)\;
      \Return\;
    }
    
    \Case{$n \in B$}{
      \texttt{Store}(n.id, n)\;
      \For{$c \in C$}{
        \FPer{c}\;
      }
      \Return\;
    }
    
    \Other{
       \texttt{Store}(n.id, n)\;
       \FPer{n.c}\;
      \Return
    }
}
}
\end{algorithm}

\begin{example}
Consider Figure~\ref{fig:livedb} and 
suppose $v_2$ and $v_4$ have been edited (indicated by the cut line outline). They will need their hashes updated, as will $n_1$ and $n_2$ and $n_0$. 
Algorithm~\ref{alg:hashcomp} will compute the sub tree of nodes $\{n_0, n_1, n_2, v_2, v_4\}$ and then compute the hashes starting with the leaf nodes $v_2$ and $v_4$ the branch nodes $n_1$ and $n_2$ and finally the root node $n_0$.
\end{example}

\subsection{On-Disk Persistent Representation}
The persistence layer implements a flattened representation of the \textsc{LiveDB} on the underlying file system, storing consecutive  nodes in binary files.
Both in-memory and disk-resident nodes reference their child nodes via a simple record number ($id$). 
The index uses 6 bytes, which is much more space conservative than original hash-based addressing, where the hashes were 32 bytes.
Each newly appended node is assigned a unique, sequential identifier, which directly corresponds to its position within the file.
This approach ensures dense storage and optimal utilization of space.

Nodes are organized into distinct files based on their type: inner, branch, leaf, reflecting the fact that each node type has a fixed and unique size.
This separation allows for efficient file compaction.
In our particular implementation, the trie includes further specialized node types to represent account states or storage slots to fit for their specific sizes and hashes are stored in dedicated files as well.
Efficient swapping between memory and disk is facilitated by direct indexing: the file system's seek operation is used to locate the node as its position on the file system is equivalent to its index times the fixed size of the node.
Furthermore, once a node is written to disk, its position remains fixed for the duration of its existence, thereby simplifying access semantics and enabling an overhead-free schema-less model.

The persistence layer is modeled as consecutive nodes in binary files: \newline $\langle \mathcal{F}, \texttt{GetID}, \texttt{Store}, \texttt{Load}, \texttt{Delete} \rangle$ where $\mathcal{F} = \mathbb{N} \rightarrow V$ is a file mapped array with the following operations: $\texttt{GetID}: () \rightarrow \mathbb{N}$, $\texttt{Store}: \mathbb{N}  \times V  \rightarrow ()$,
$\texttt{Load}: \mathbb{N}  \rightarrow V$,  $\texttt{Delete}: \mathbb{N}  \rightarrow ()$. The operators are described in Algorithm~\ref{alg:player}.  \texttt{GetID} picks an $id$ from the $freelist$ of $id$s or if there are no free $id$s, generates a new $id$. \texttt{Store} stores a fixed size serialised node in the position of the $id$. The nodes are serialised by converting all its attributes into a consecutive byte array. All attributes are fixed size, which produces a fixed size array. The key of leaf and inner node is extended to a zero padded 32B array to make it the fixed size.   
\texttt{Load} loads a serialised node from the position of the $id$. \texttt{Delete} deletes a node under the $id$ and adds the $id$ to the $freelist$ for later reuse. Although this may temporarily introduce fragmentation and unused space—seemingly at odds with the principle of dense storage—newly created nodes rapidly reuse vacant positions. The system prioritizes placement of new nodes into available free slots rather than appending them to the end of the file.

\SetKwFunction{FGetID}{\texttt{GetID}} 
\SetKwFunction{FStore}{\texttt{Store}} 
\SetKwFunction{FLoad}{\texttt{Load}} 
\SetKwFunction{FDelete}{\texttt{Delete}} 

\begin{algorithm}
\caption{Persistence Layer Operators\label{alg:player}}
\KwData{$freelist$}
\KwData{$\mathcal{F}$}

\Fn{\FGetID{}}{
  \If{$freelist = []$}{
    $id \gets generate()$\;
    \Return $id$
  } 

  $id \gets head(freelist)$\;
  $freelist \gets tail(freelist)$\;
  \Return $id$\;
}

\Fn{\FStore{id, v}}{
$\mathcal{F}$[id] $\gets$ v
}

\Fn{\FLoad{id}}{
  \Return $\mathcal{F}$[id]
}

\Fn{\FDelete{id}}{
$freelist \gets id::freelist$\;
$\mathcal{F} \gets \mathcal{F} - (id, \mathcal{F}[id])$\;
}
\end{algorithm}

The hashes can be recomputed in-memory only. For this, the nodes must be obtained from the disk first, and they cannot be persisted back without computing their hashes. Practically, the hashes are computed at the end of each block, and for this reason, memory has to accommodate enough nodes for the completion of one block. For performance reasons, the memory is not necessarily cleaned after each block; however, the nodes are kept using the Least Recently Used (LRU) cache policy, as nodes close to the root are often needed repeatedly.

Updates of values in the tree and updates of the hashes are two distinct operations that are orthogonal to each other. Since hashing is an expensive operation; it can only be executed when the Virtual Machine (VM) and Block Processor require the root hash, i.e., it is executed only on demand. It can even be performed asynchronously to block processing.

\begin{figure*}
\centering
\begin{subfigure}{0.45\textwidth}
\includegraphics[width=0.95\textwidth]{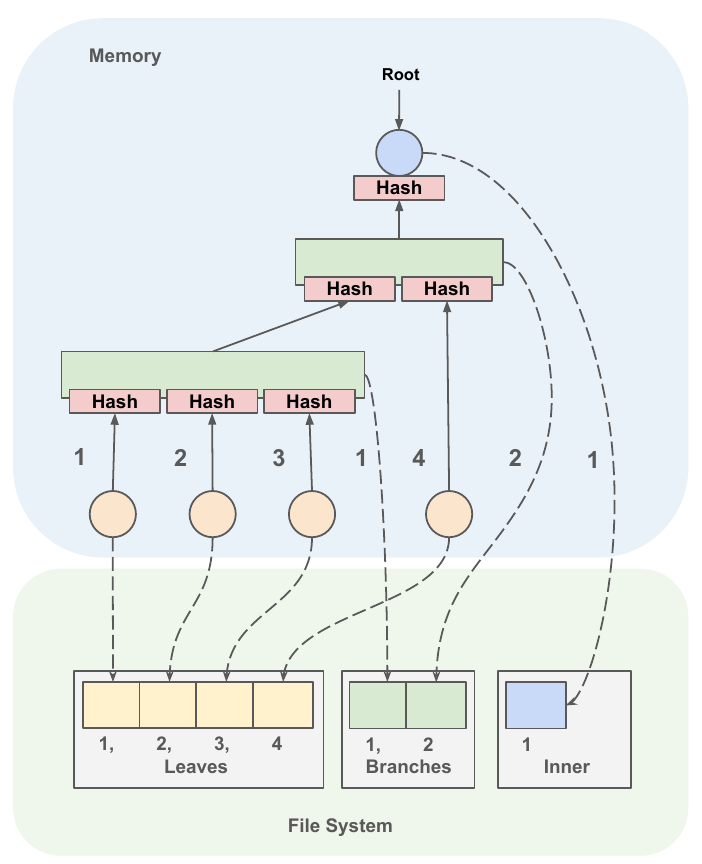}
\caption{Before Update}
\label{fig:live-before}
\end{subfigure}
\hfill
\begin{subfigure}{0.45\textwidth}
\includegraphics[width=0.95\textwidth]{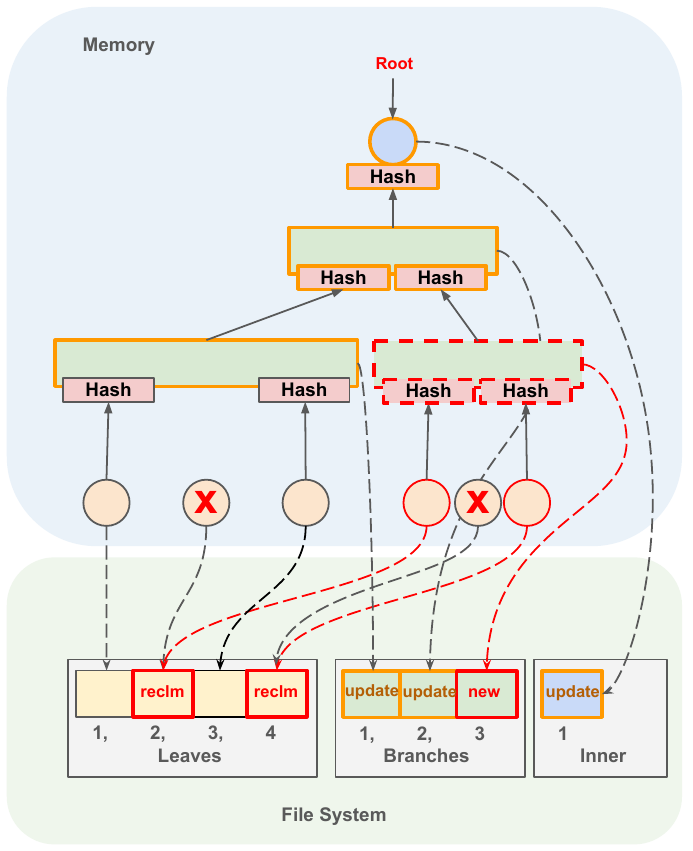}
\caption{After Update}
\label{fig:live-after}
\end{subfigure}
\caption{LiveDB and its Persistance}
\label{fig:live-all}
\end{figure*}

\begin{example}
Consider Figure \ref{fig:live-all}. It provides a visual representation of the principles discussed above. Specifically, Figure \ref{fig:live-before} illustrates an \textsc{LiveDB} in a particular state, comprising one inner node (blue), two branch nodes (green), and four leaf nodes (yellow). The upper portion of the figure (blue area) corresponds to the in-memory representation of the \textsc{LiveDB}, while the lower portion (green area) depicts its disk-resident counterpart. Each node references its children via a numeric pointer, which directly maps to the child node’s index within the binary file. For clarity, the figure includes connecting lines between nodes in the in-memory layer to convey the trie structure visually; however, in practice, these connections are realized solely through positional indices in the underlying storage files.

The Figure \ref{fig:live-after} on the right illustrates an update to the structure. It shows addition of a few new nodes marked by red, and removal of some nodes, marked by a red cross. Specifically, a leaf node was removed from the left-most branch, and a new leaf was added to the right-most branch. This addition required the creation of an additional branch node, with new nodes depicted in red. As \textsc{LiveDB} is an updatable structure, the corresponding binary files are updated as follows: one position (index 2) is freed from the file but is subsequently reclaimed by the newly created leaf added to the right subtree. Additionally, a leaf node on the right was replaced by a different node due to the update, and its position in the binary file was similarly reclaimed (index 4), and one more branch node is added (index 3). Following this operation, the remaining branch and inner nodes must be updated to reflect the new indexes of child nodes, and their hashes must be recalculated accordingly. These changes occur in-place, meaning that the corresponding nodes in the binary files are updated and persisted.  The updated nodes are highlighted in yellow in the figure.  
\end{example}


\subsection{Discussion}
In this subsection we discuss the rational of the design the \textsc{LiveDB} and contrast it to existing Ethereum 
state-of-the-art databases.


\paragraph{Intrinsic Pruning}
A core feature of the \textsc{LiveDB} is its mutable tree structure. We have opted for mutability to avoid the need for pruning and unnecessary copying of data. This contrasts to the Ethereum stateDB which is based on the append only MPT~\cite{mpt} data structure. Due to its append only semantics, an update of a value in a leaf nodes leads to the creation of a new leaf node and its hash value. This triggers a cascading effect and all nodes on the path from the root to the leaf node need to be cloned as a result of an update parent-child relation. In the case of validator/observer nodes in a forkless blockchain, this requires costly \emph{pruning}. The pruning effort is substantial because it needs to traverse the whole tree from its root, discovering all nodes that belong to the tree. Only then, the remaining nodes may be trashed. To overcome the pruning problem associated with continuous updates of values (such as account balances, nonce, and storage keys), the \textsc{LiveDB} avoids creating a new leaf node, instead modifies the leaf node for the given key and mark the leaf as dirty. This is performed for all updates in a block resulting in potentially several leaf nodes marked as dirty by setting a \emph{dirty} flag as true. Since the nodes along the path of the leaf nodes  are no longer referenced by their hash and instead use their index numbers, intrinsic pruning can be performed without altering the tree structure.   

\begin{example}

\begin{figure*}[htbp]
        \centering        \includegraphics[width=0.55\textwidth]{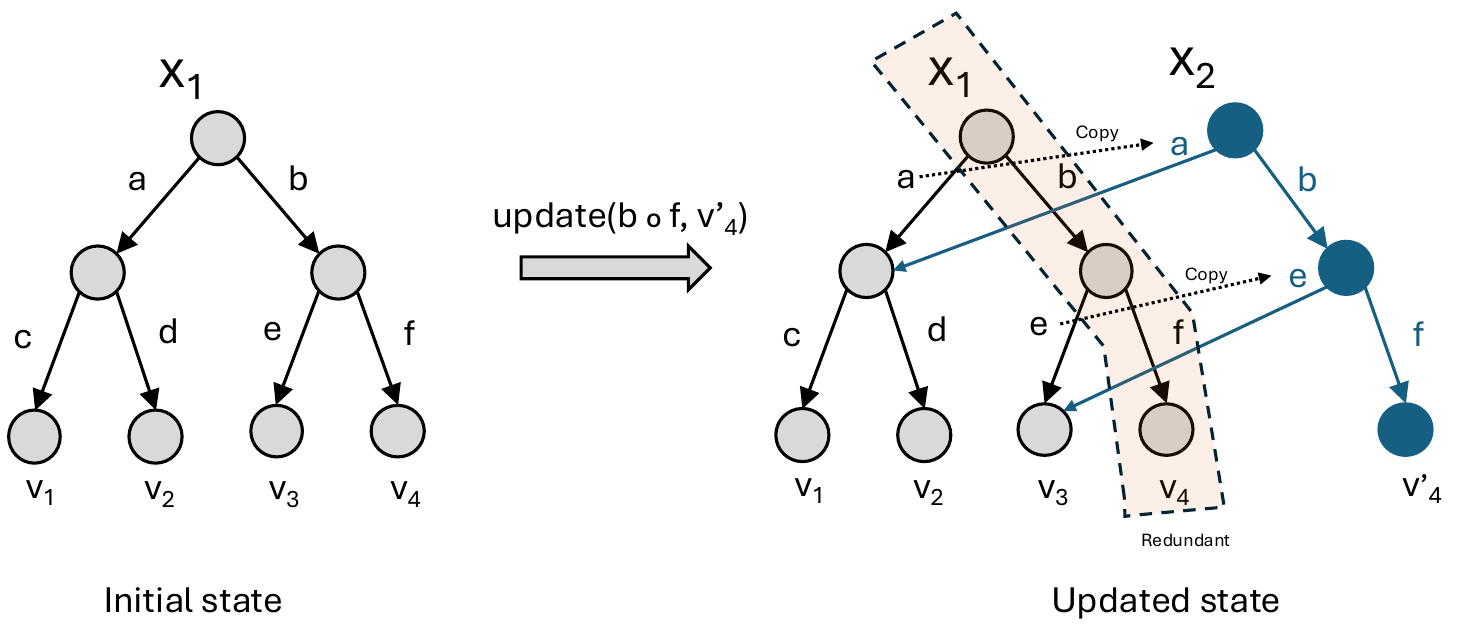}
        \vspace{-3mm}
        \caption{Evolution of Ethereum stateDB: A new version of state $x_1$ is constructed, resulting in a new state $x_2$ by updating the value $v_4$ to $v_4'$ for the key $b \cdot f$. \label{fig:pruning}}
\end{figure*}
Consider the MPT on the left in Figure~\ref{fig:pruning}. This represents the MPT in an initial state. It consists of a single root $x_1$ that describes the following key $k_i$ to value node mapping: $k_1 \mapsto v_1, \ldots, k_4 \mapsto v_4$ where $k_1 = a \cdot c$, $k_2 = a \cdot d$, $k_3= b \cdot e$, and $k_4 = b \cdot f$. If we execute a transaction, the mapping changes from $k_4 \mapsto v'_4$ as shown on the right-hand-side of Figure~\ref{fig:pruning}. Here we introduce a new root $x_2$ representing the new version and duplicate the path along edges $b$ and $f$. The highlighted path for a forkless blockchain is redundant, as it is not needed for the new root $x_2$. Retaining the data causes bloat and thus will eventually be \emph{pruned}~\cite{livepruning}, which typically requires putting the node offline.  Moreover, as shown by the dotted arrows, numerous copies of keys along the path need to be copied to create a path from $x_2$ to the new value $v'_4$. The mutable tree design of the \textsc{LiveDB} avoids this problem. 
\end{example}

\paragraph{File Mapped Array Persistence Layer}


Compared to the Ethereum stateDB, the \textsc{LiveDB}
avoids the use of a key-value store~\cite{rocksdb, leveldb} as a storage layer.  The use of a key-value database in the 
Ethereum stateDB leads to a phenomenon known as \emph{read amplification}.  In this setup, the key-value store becomes a pile of hash-node pairs, where the node of the MPT is encoded as an RLP string, and the key is the hash of the RLP-encoded string. When, for example, the balance of an account is accessed in the MPT, we have to load each node into memory (assuming the node is not already in a cache), exhibiting a worst-case complexity of ${\mathcal O}(log(n))$ loading a single node where $n$ is the number of keys in the key-value storage. Since it is a trie, we have ${\mathcal O}(log(n))$ inner nodes for an attribute access, resulting in a combined worst-case complexity of ${\mathcal O}(\log^2 n)$. The question is whether we can achieve a better access complexity for MPT nodes. To mitigate these limitations associated with hash-based node addressing, \textsc{LiveDB} uses an alternative design in which cryptographic hashes are decoupled from the addressing mechanism. In this model, each node retains its hash as a secondary attribute for integrity verification, while child references are implemented using direct pointers—specifically, ordinal indices. This enables constant-time access, as nodes can be stored in a contiguous array and referenced by index values in the range $[0..N]$,  resulting in lookup complexity of ${\mathcal O}(1)$.

\begin{figure*}
\centering
\begin{subfigure}{0.45\textwidth}
\includegraphics[width=0.95\textwidth]{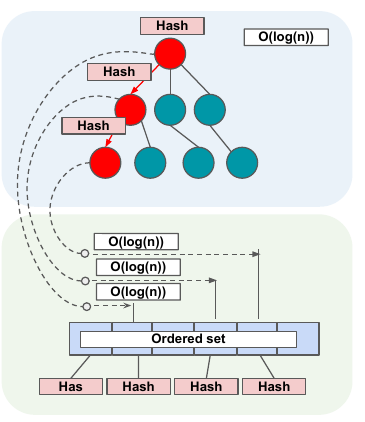}
\caption{Loopup O(log(n)) for Hash Keys}
\label{fig:keys-as-hashes}
\end{subfigure}
\hfill
\begin{subfigure}{0.45\textwidth}
\includegraphics[width=0.95\textwidth]{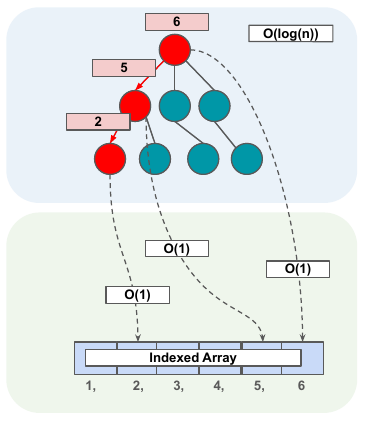}
\caption{Lookup O(1) for Index Keys}
\label{fig:keys-as-indices}
\end{subfigure}
\caption{Distinction Between Nodes Addressed as Hashes or as Indices with its Impact on Access Time Complexity}
\label{fig:key-addresses}
\end{figure*}

\begin{example}
The principle is illustrated in Figure \ref{fig:key-addresses}. The left panel depicts the conventional approach, wherein each node is retrieved from the underlying storage using its hash. This retrieval typically involves a search operation with worst-case complexity of ${\mathcal O}(\log n)$, assuming the storage engine employs a sorted structure or auxiliary indexing mechanism. In contrast, the right panel demonstrates the proposed method, where nodes are addressed by their indices. In this configuration, each node can be directly accessed from an array structure, eliminating the need for costly search operations and enabling efficient traversal.
\end{example}

\paragraph{Separation of Hash and ID}
A drawback of the MPT is the redundant storage of hashes in the key-value store. It uses the hashes to refer to child nodes, and these references are stored with parent nodes. Taking into account that a branch node has 16 references to children, and the hash size is 32 bytes, one branch node needs half a kilobyte to refer to all the children. However, when a child node is updated, its new hash is recomputed and stored within the parent. The parent node is also recomputed, creating new nodes from the bottom up. In the worst case, only 32 bytes are updated for each node along this path, while half a kilobyte is copied for each node. This leads to a bloat of hashes in the database.  

\begin{figure*}
\centering
\begin{subfigure}{0.45\textwidth}
\includegraphics[width=0.95\textwidth]{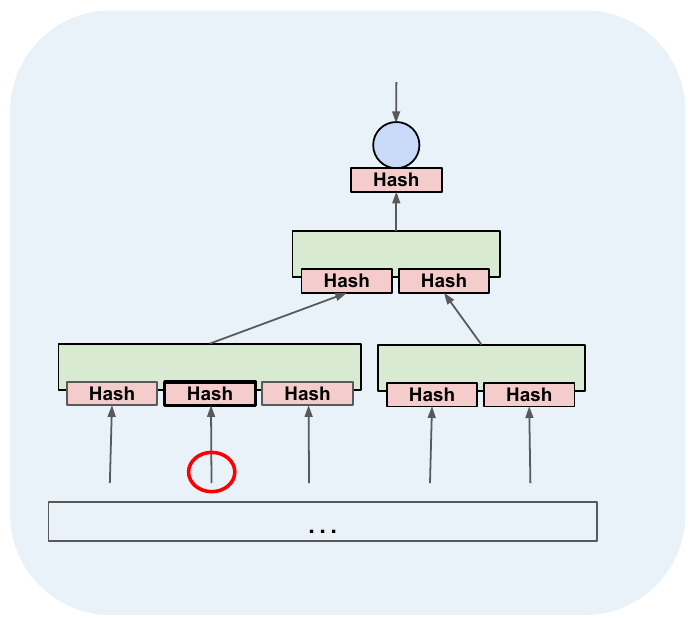}
\caption{Update in Child Node }
\label{fig:hash-update}
\end{subfigure}
\hfill
\begin{subfigure}{0.45\textwidth}
\includegraphics[width=0.95\textwidth]{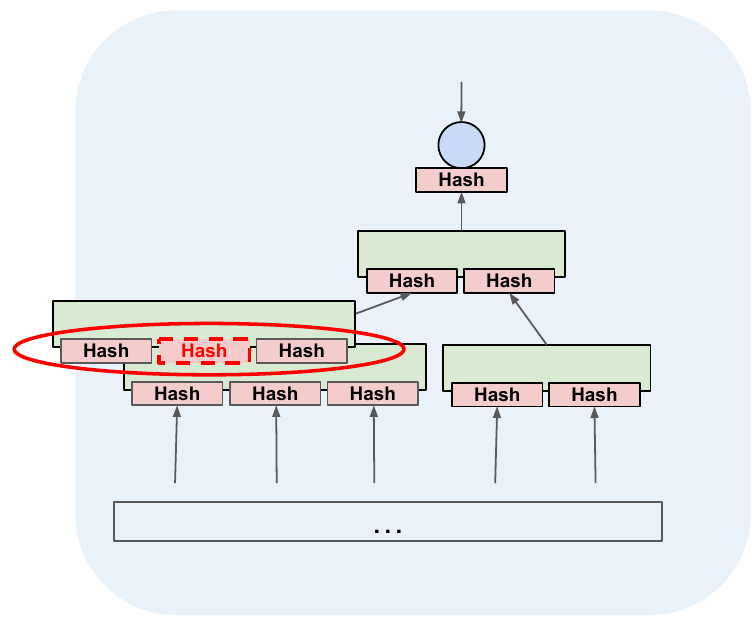}
\caption{Parent Node is Copied}
\label{fig:hash-update-copy}
\end{subfigure}
\caption{Update in child node triggers hash calculation that must copy the whole parent node with all other hashes}
\label{fig:hash-copy-update-write}
\end{figure*}

\begin{example}
This principle is demonstrated in Figure \ref{fig:hash-copy-update-write}, showing that each update triggers copying the whole nodes. When the history is kept, and no expensive pruning is used, the database becomes bloated with many copies of the same hashes.
\end{example}

The \textsc{LiveDB}  deliberately stores node hashes within parent nodes to facilitate rapid recomputation following structural changes rather than having the hash as an attribute inside the node.
This is because the hash of each node is deterministically derived from the hashes of its child nodes, where the computation proceeds recursively in a bottom-up fashion, beginning at the modified node and propagating upward to the root. Since branch nodes in the MPT maintain 16 child references, each associated with a corresponding hash, the recomputation of a branch node’s hash necessitates access to all 16 child hashes. To optimize this process, the system embeds child hashes directly within the parent node. This design ensures that, upon modification of a child node, the parent node can immediately recompute its hash without incurring additional disk reads of all siblings. 

\begin{figure} 
\centering 
\includegraphics[width=0.5\textwidth]{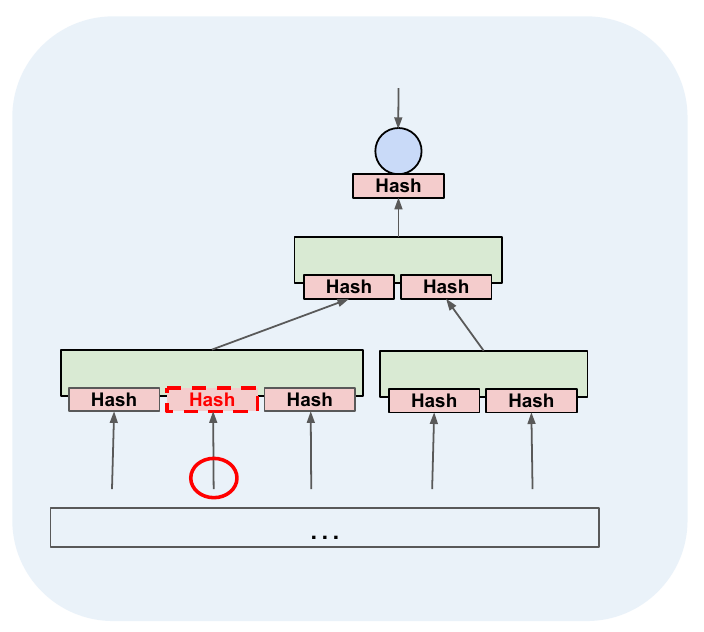} 
\caption{In-place Updates in LiveDB} 
\label{fig:in-place-update} 
\end{figure}

\begin{example}
    This principle is depicted in Figure \ref{fig:in-place-update} where each parent node contains hashes of children. When a child is modified, it happens in-place in this child, and the hash in the parent is also modified in-place. In particular, no node is copied. When the parent node is re-hashed, it includes the modified hash, while other sibling node hashes are already available within this node.  
\end{example}

Despite counterintuitive, the hash storage strategy employed by LiveDB exhibits inherent space efficiency. \textsc{LiveDB}  is architected as an updatable data structure that supports destructive updates, allowing for direct in-place modifications of existing values in the MPT.
Upon updating a value, the corresponding node is located and overwritten. Deletion operations remove the associated nodes, while insertions result in the creation of new nodes. This update mechanism ensures that the MPT maintained by \textsc{LiveDB}  consistently retains a strict tree structure. Each child node is referenced by a single parent, as nodes are never duplicated upon update. This design prevents scenarios in which multiple parent nodes point to the same child, a condition that would otherwise transform the tree into a directed acyclic graph (DAG).

In contrast, the original MPT design adopts a copy-on-write strategy, as illustrated in Figure \ref{fig:hash-copy-update-write}, where in any modification to a node results in the creation of a new version.  This leads to the situation where the MPT becomes DAG. Since multiple parent nodes then reference the same child node via identical 32-byte hashes, each parent must store an identical hash to maintain the reference in the hash based reference schema. These duplications demand extensive storage allocations. 

The operational model of our \textsc{LiveDB} ensures a strict one-to-one correspondence between the number of nodes and the number of stored hashes. Even when the hashes are stored with parents, there is no redundancy as in the tree structure each child has exactly one parent. 
For this reason, the system maintains exactly one hash per node, thought not stored together. As a result, the total hash footprint mirrors the node count, achieving optimal space utilization.

\paragraph{Ethereum Compatibility}

\textsc{LiveDB} supports the extraction of supplementary information, including witness proofs and state root hash computations that are compatible with Ethereum. This capability is particularly advantageous for enabling interoperability across blockchain networks, as it allows data stored on one chain to be directly compared and verified against data from other chains that use Ethereum's implementation. 

While our design introduces a novel storage layout and hash representation that diverges from the classical Ethereum architecture, the in-memory structure of the trie remains consistent with its legacy form. Specifically, it retains the original node types and encodes key-value pairs using the same logical scheme, including the use of RLP encoding for node representation.

Unlike the legacy approach, in which RLP-encoded nodes are persisted directly in the database, our design serializes nodes as plain byte strings with attributes laid out in a canonical format. This strategy reduces storage overhead, as RLP encoding includes explicit length descriptors for each attribute—an inefficiency in contexts where node attributes are of fixed size and their lengths need not be stored. In our system, RLP encoding is applied only during in-memory representation, solely for hash computation. For consistency with existing cryptographic practices, we employ the Keccak \cite{bertoni2013keccak} hashing function to derive node hashes from their RLP-encoded forms.


\section{ArchiveDB}
\label{sec:archivedb}


\textsc{ArchiveDB} is a dedicated database, separated from \textsc{LiveDB}, designed to support historical queries for clients requiring access to prior blockchain states. Given the scale of modern blockchains—often requiring terabytes of storage—and the limitations of conventional databases, particularly with respect to read amplification, we extend the techniques employed in \textsc{LiveDB} to \textsc{ArchiveDB}, with modifications that preserve historical state. Specifically, destructive updates are permitted, but only for nodes that are not part of any previously committed state. Whereas \textsc{LiveDB} aggressively prunes obsolete state across transaction and block boundaries to maintain minimal footprint and high throughput, \textsc{ArchiveDB} must retain historical snapshots. To achieve this, pruning is applied selectively, and at each block boundary, the global state is partially frozen to ensure immutability for archival purposes. These frozen states remain accessible for retrospective analysis and are insulated from subsequent transaction-induced modifications.

Standard database means for resiliency, which include journaling and a system of transactions that can be atomically committed or rolled back, are expensive in the context of blockchain. 
The reason is that they employ backup structures that require copying the data back and forth.
This does not apply to fast file-mapped arrays. 
For this reason, we propose a mechanism where data consistency is guaranteed only at selected checkpoints, and users are required to rollback the database to the latest checkpoint should an error occur.
This assumes that data above the checkpoint can be lost as long as the checkpoint is relatively recent, and re-syncing the missing data is not expensive. 

Finally, the representation of hashes has different requirements for \textsc{ArchiveDB}. While the demand on \textsc{LiveDB} is for the fast recomputation of hashes to evolve the blockchain tip, \textsc{ArchiveDB}  must be space-conservative. 
Since the hashes on the archive are never recomputed again, but their number grows with the number of nodes, the Archive must be designed to minimize disk usage. 


\subsection{Read Amplification and Pruning}
\textsc{ArchiveDB} adopts a design paradigm fundamentally aligned with that of \textsc{LiveDB}, wherein the tree is maintained in-memory, and data exceeding memory capacity is offloaded to binary files. 
The principal difference between the two systems lies in ArchiveDB’s handling of multiple versions of the world state and its approach to making the world state immutable.
Unlike \textsc{LiveDB}, which permits destructive, in-place updates, \textsc{ArchiveDB} enforces a non-destructive update strategy at a block boundary by freezing the world state of the preceding block.

In \textsc{ArchiveDB}, state updates are governed by a copy-on-write protocol designed to preserve historical integrity. When a node requires modification, a new version is created by copying and altering the original, which is then stored independently. This procedure is applied recursively along the traversal path from the root to the target node, ensuring that all ancestor nodes reflect the updated lineage while maintaining immutability of prior versions. Each node in \textsc{ArchiveDB} carries a frozen flag that dictates its update semantics: frozen nodes are subject to copy-on-write, while non-frozen nodes may be modified in-place, mirroring the behavior of \textsc{LiveDB}. This mechanism is central to block-level version lifecycle management. During block execution, multiple transactions may interact with the same nodes; however, only the final state at the end of the block is considered canonical. To enforce this, the system seals the block by marking all updated nodes as frozen. Any subsequent modification to these sealed nodes in future blocks triggers the copy-on-write protocol, thereby preserving historical consistency.

\begin{example}
\begin{figure*}
\centering
\begin{subfigure}{0.45\textwidth}
\includegraphics[width=0.95\textwidth]{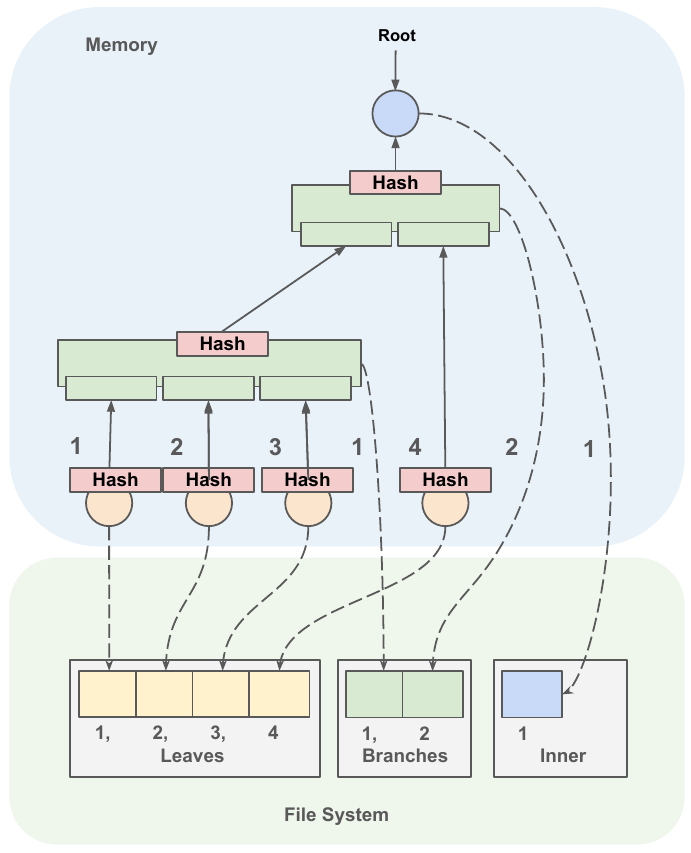}
\caption{Before Update}
\label{fig:archive-before}
\end{subfigure}
\hfill
\begin{subfigure}{0.45\textwidth}
\includegraphics[width=0.95\textwidth]{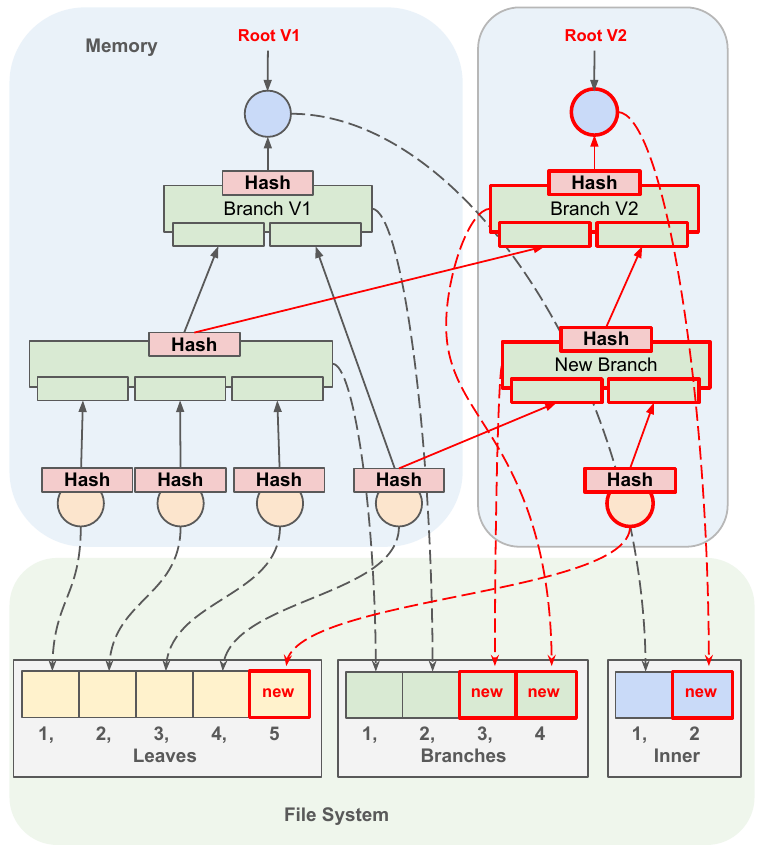}
\caption{After Update}
\label{fig:archive-after}
\end{subfigure}
\caption{\textsc{ArchiveDB} and its Persistence}
\label{fig:archive-all}
\end{figure*}

Figure \ref{fig:archive-all} demonstrates the process of updating the Archive database.
The left-hand image (Figure \ref{fig:archive-before}) depicts a structure similar to LiveDB, with the primary difference being the method of hash storage.
However, the update process diverges significantly.
Assume that all nodes are already frozen, and the tree is updated in a manner that necessitates the addition of a new leaf node, as illustrated in Figure \ref{fig:archive-after}.
This addition consequently requires the creation of a new branch node (highlighted in red), to which both the previous and new leaf nodes are attached.
Given that the Archive database operates as an append-only structure, this new branch node cannot be attached to the previous parent.
Instead, the previous parent (a branch node, denoted as V1) is copied, and the new branch node is attached to this copy. Furthermore, the parent of the original branch, a root extension node, must also be copied to serve as the new root.
This design enables clients to query any version of the tree, ensuring that they consistently access the correct nodes and values. 
\end{example}

\subsection{Hashing}
\textsc{ArchiveDB} adopts a fundamentally different strategy for hash storage compared to LiveDB.
In \textsc{LiveDB}, hashes are embedded within parent nodes to enable rapid recomputation across the trie.
Conversely, \textsc{ArchiveDB} stores hashes directly within child nodes, a design motivated by disk space optimization.
\textsc{ArchiveDB} copies the nodes upon modification. This makes the original tree a DAG where one child can be referred by more than one parents. 

Storing child hashes within parent nodes introduces significant space inefficiencies when used in the \textsc{ArchiveDB}. In particular, in the case of branch nodes, which may reference up to 16 children and thus require storage of 16 distinct 32-byte hashes. This inefficiency is exacerbated by the fact that the trie structure effectively forms a directed acyclic graph (DAG), wherein a single child node may be referenced by multiple parents. If each parent redundantly stores the same child hash, the cumulative overhead becomes substantial. In contrast, embedding the hash directly within the child node—rather than duplicating it across all referencing parents—eliminates redundancy and yields a more space-efficient representation. This design ensures that each node encapsulates its own cryptographic identity, thereby avoiding duplication.

\begin{figure*}
\centering
\begin{subfigure}{0.45\textwidth}
\includegraphics[width=0.95\textwidth]{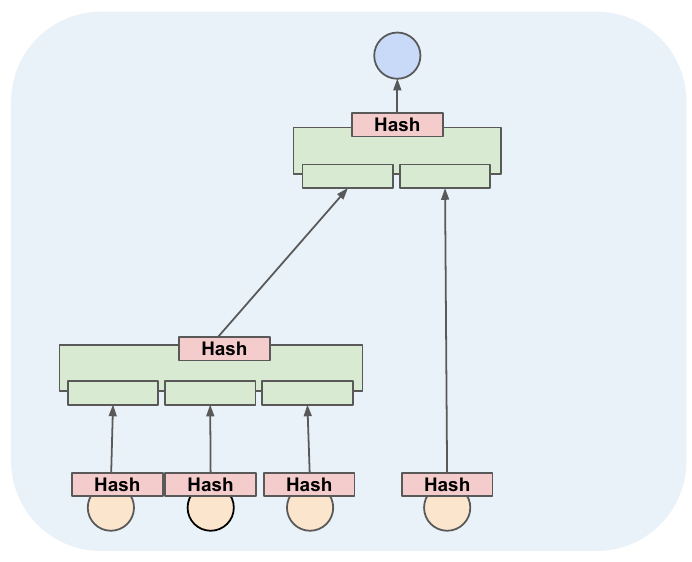}
\caption{Update in a parent node}
\label{fig:archive-update-before}
\end{subfigure}
\hfill
\begin{subfigure}{0.45\textwidth}
\includegraphics[width=0.95\textwidth]{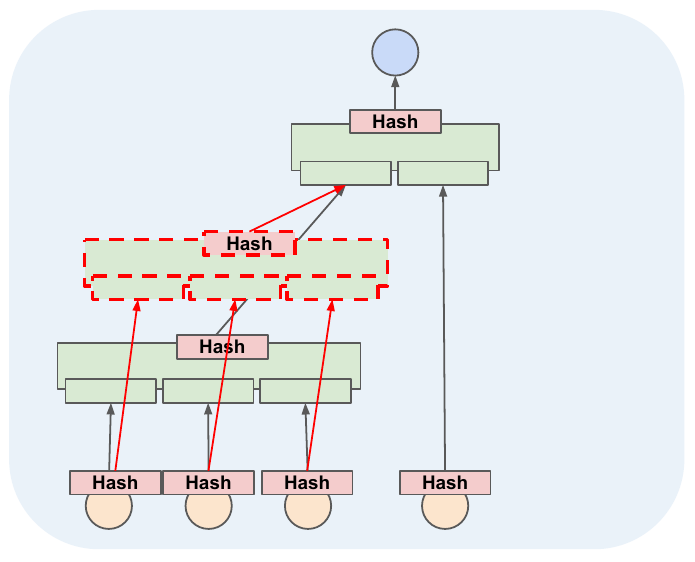}
\caption{Only the updated node and its own hash are copied}
\label{fig:archive-update-after}
\end{subfigure}
\caption{\textsc{ArchiveDB} update operation}
\label{fig:archive-update}
\end{figure*}

\begin{example}
This principle is depicted in Figure \ref{fig:archive-update} where one of the nodes is modified. Since it is a parent node of other nodes, the copy of this node makes the children referring to two parents, and the tree becomes a DAG. However, as shown in the right panel \ref{fig:archive-update-after} only the affected node with its own hash is copied. For this reason, no redundancies are created as it is the case in the original solution presented in Figure \ref{fig:hash-copy-update-write}, where child nodes are stored with the parent. 
\end{example}

Since \textsc{ArchiveDB} stores hashes directly in child nodes, recomputing a branch node’s hash would require explicitly loading the hashes of all 16 child nodes from the database, unless they are already cached, resulting in up to 16 individual read operations per branch.
Although historical hashes in \textsc{ArchiveDB} are not updated, they must be recomputed when new data is appended, specifically during the propagation of the latest \textsc{LiveDB} state to \textsc{ArchiveDB} at the end of each block.
To mitigate the overhead of recomputation, we employ an optimization: instead of recalculating hashes within \textsc{ArchiveDB}, we copy the already computed hashes from \textsc{LiveDB}.
This approach ensures that no additional hash computation occurs in the archive, thereby preventing the potential performance issue from manifesting.

\subsection{Healing}

The proposed design adopts a performance-oriented strategy that deliberately relaxes traditional consistency guarantees.
Unlike conventional ACID-compliant systems, the database writes directly to binary files without supporting rollback mechanisms or providing explicit confirmation of successful data persistence.
Instead, it relies on error codes returned by the underlying file system to detect potential I/O failures.
In the event of such failures, the database is deemed corrupted and must undergo a recovery process before it can resume operation.

ArchiveDB’s recovery mechanism is structured around a system of checkpoints. At designated intervals, specifically when no errors have been detected, the database records a checkpoint corresponding to the current timestamp. Each checkpoint serves as a marker of consistency, signifying that all data up to that moment is considered valid and intact. If a failure occurs, ongoing operations are halted to prevent further corruption, and the user is required to restore the database to the most recent valid checkpoint, thereby ensuring a consistent recovery state.

This design choice is motivated by the substantial overhead associated with traditional database systems, which are grounded in the principles of atomicity, consistency, isolation, and durability—collectively known as the ACID properties. These guarantees ensure data integrity across transactions but come at a high computational and storage cost. Conventional databases rely on mechanisms such as transactions, commit protocols, rollbacks, and journaling to uphold ACID compliance. Typically, data is first recorded in a journal at the initiation of a transaction; upon commit, the journaled data is propagated to persistent storage, while rollback operations discard the changes.

While this approach may appear disruptive—requiring users to pause standard operations during healing—the trade-off yields significantly improved performance under normal conditions. This design philosophy is motivated by the Pareto Principle \cite{pareto1896cours}, which suggests that approximately 80\% of system runtime proceeds without incident, while only 20\% involves recovery procedures. Thus, users predominantly benefit from the system’s high-throughput behavior, with infrequent interruptions representing a smaller operational cost.

From a technical standpoint, checkpoints are implemented at the level of binary files where the nodes are persisted.
Since the tree structure resides only temporarily in memory, recovery does not necessitate reconstructing the memory.
Instead, consistency is ensured through the management of the underlying binary files. These files may be extended via appends or may contain reclaimable space internally. 
Checkpoint creation is initiated by flushing all buffered data to disk and invalidating the free list. This process relies on the file system’s guarantee that a successful return from the flush operation implies durable persistence of the written data. In cases where data is appended to the end of a file, flushing ensures that all additions are committed to disk.

To prevent inconsistencies, any previously available free slots within the file are abandoned by clearing the free list. This measure ensures that sealed files are not inadvertently modified by reusing internal slots.
Parallel writes to the end of the file are permissible, as such operations merely append data that can be safely ignored during recovery. Crucially, no data is written within the sealed region of the file, preserving its integrity. The resulting file state is thus considered consistent. A trade-off of this approach is the potential loss of reclaimable space due to the invalidation of the free list during checkpoint creation.

Due to the distributed nature of data storage across multiple binary files, checkpoint creation is implemented as a two-phase protocol, inspired by the classical two-phase commit mechanism \cite{gray1978notes}. In the first phase, a designated coordinator initiates the checkpoint process by requesting all participating files to prepare for checkpointing. Each file responds by flushing its buffered data to disk and reporting the success or failure of the operation. If all participants confirm successful flushing, the coordinator proceeds to the second phase, issuing a commit signal that instructs each file to record the new checkpoint. Conversely, if any participant fails to signal success, the checkpoint creation is aborted and deferred for a subsequent attempt. 

\begin{figure}
\centering
\includegraphics[width=0.3\textwidth]{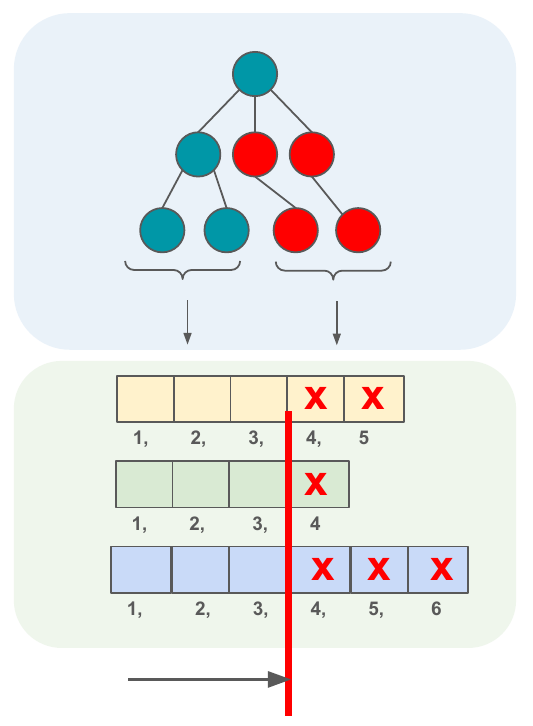}
\caption{Recovery From Checkpoint}
\label{fig:checkpoint}
\end{figure}

\begin{example}
Given the checkpointing system described above, the recovery mechanism is conceptually straightforward and illustrated in Figure \ref{fig:checkpoint}. Upon detection of corruption or failure, the healing process selects the most recent successful checkpoint and truncates all binary files to that point. Any data written beyond the checkpoint is discarded.
Recovery is performed while the system is offline, i.e., when no in-memory representation is active, and the binary files are not being modified. Upon system restart, the in-memory model is reconstructed using only the nodes covered by the checkpoint; nodes beyond this boundary are excluded.  
\end{example}

Because checkpoints are tied to specific timestamps, all data preceding the checkpoint must remain available, whereas data written afterward can be safely abandoned.
This recovery strategy leverages the append-only nature of ArchiveDB’s binary files, making it applicable exclusively to archival storage, where data is never overwritten or deleted. In contrast, LiveDB requires a separate recovery path: the operator must first restore ArchiveDB, from which LiveDB can be reconstructed using the most recent valid blocks.

It is also important to note that, due to the decentralized architecture of blockchain systems, even in the event of complete checkpoint failure, recovery remains feasible.
The system can resynchronize its state by retrieving data from peer nodes in the network.


\section{Optimizations}
\label{sec:opt}
While the redesigned architecture already yielded notable performance gains by eliminating read amplification, additional optimizations were implemented to accelerate the processing of real-world blockchain datasets. These enhancements were informed by profiling the initial prototype against actual blockchain data, wherein CPU usage patterns were analyzed and computational hot spots identified.

To mitigate these bottlenecks, a multi-tiered caching strategy is introduced. First, a memory-resident subset of the nodes establishes a node-level cache. This cache adheres to a Least Recently Used (LRU) eviction policy, ensuring that frequently accessed nodes, particularly those close to the root, remain available in memory most of the time. Notably, this cache constitutes the largest memory footprint among the caching layers.

Second, a dedicated State cache is incorporated to store key-value pairs directly. Although the database inherently encodes key-value mappings, retrieving them via the node cache necessitates traversing the trie from root to leaf, incurring an access complexity of  ${\mathcal O}(\log n)$. In contrast, the State cache enables constant-time retrieval ${\mathcal O}(1)$ for frequently accessed keys, thereby reducing lookup latency.

Although the revised design achieves efficient file system utilization through dense data storage and the elimination of costly reorganization procedures, I/O operations remain a performance hot spot. To mitigate this, we introduce an intermediate buffering mechanism between the node cache and the files. Specifically, when a node is evicted from the cache, it is not written to disk immediately. Instead, it is temporarily held in the buffer that aggregates multiple evicted nodes.
This buffer is periodically flushed to disk in batches by a dedicated parallel thread, thereby reducing the frequency and overhead of individual write operations. To further enhance write locality and minimize disk seek times, nodes within the buffer are sorted by their respective file-system indexes prior to flushing. This ordering increases the likelihood that nodes are written contiguously, improving overall I/O throughput.

Given that the node cache is in practical usage huge, it presents an operational challenge. Prior to client termination, the cache must be propagated to disk, a process that may take several seconds or even minutes on resource constrained devices. Such latency is risky in scenarios where a human operator or containerized orchestration system may interpret the delay as a hang and forcibly terminate the process. This premature termination leads always to database corruption, necessitating  the healing procedures.
To address this issue, we implemented a periodic background flushing strategy. A dedicated thread asynchronously flushes the contents of the node cache to disk at regular intervals, thereby minimizing the volume of unpersisted data at any given time. As a result, the cache typically contains only a small number of transient nodes, significantly reducing shutdown latency.

To address structural imbalances inherited from original Ethereum's design, we introduce a dedicated thread for account deletion. The database logically partitions data between accounts and their associated storage, with each storage subtree linked to a specific account. In practical blockchain deployments, this results in significant asymmetry: some accounts maintain extensive storage structures, while others contain minimal data. Consequently, the trie becomes unbalanced.
A performance and reliability concern arises when deleting accounts with large storage footprints. Due to the design of our database, the nodes can be refereed only by their indexes, and they must be processed individually, the deletion procedure involves traversing and marking each node for removal on the file system. For accounts with substantial storage, this operation can be time-intensive and may expose the system to denial-of-service (DoS) vulnerabilities if triggered on purpose.
To mitigate this risk, we adopt a hybrid deletion strategy. The account metadata is removed synchronously within the main application thread, ensuring immediate logical consistency. In contrast, the associated storage subtree is deleted asynchronously via a background thread.

\section{Evaluation}
\label{sec:eval}
We conducted a comprehensive evaluation to assess both transaction throughput and disk space utilization across multiple implementations of the proposed system. Our measurements focused on three core contributions. First, we demonstrated that reduced read amplification significantly accelerates block processing on a production-grade blockchain when the original database is replaced with our optimized design. Second, we evaluated how intrinsic pruning mechanisms in \textsc{LiveDB} lead to substantial reductions in disk space consumption compared to an unpruned baseline. Third, we showed that \textsc{ArchiveDB} achieves further disk space savings through a novel hash organization strategy, which eliminates redundancy inherent in the original database layout.

To facilitate these experiments, we extended the blockchain recording and replay framework introduced in \cite{Kim2021off-the-chain}, adapting it to capture execution data from the Sonic mainnet. The dataset used in our experiments consisted of approximately 34 million blocks, which were locally stored on high-performance NVMe SSDs. Using the modified tool, we replayed these blocks through selected subsystems of the Sonic client to reconstruct the blockchain state. Our modifications focused specifically on the block execution pipeline—comprising the Sonic Virtual Machine, and the underlying database—while deliberately excluding components such as peer-to-peer networking, consensus algorithms, and transaction pool logic. This isolation enabled precise measurement of transaction processing rates and associated storage demands.

All experiments were executed on a uniform environment, where the database backend was systematically varied. We benchmarked the original database inherited from Ethereum (geth client), leveraging LevelDB as its storage backend, and our \textsc{LiveDB} and \textsc{ArchiveDB}. The data was consistently stored on local NVMe disks, and we recorded metrics pertaining to both throughput and disk consumption.

\begin{description}\itemsep-1pt \item[\textbf{\em Claim-I: Throughput Enhancement by Removed Read Amplification.}] We assess the transaction processing capabilities of our system variants and compare them against the performance of the customized Geth Ethereum client previously deployed on the Sonic network. Since the new design vastly avoids read amplification by directly indexing a file mapped arrays, we contribute performance gain to this new strategy. 

\item[\textbf{\em Claim-II: Storage Efficiency of Pruning.}] We evaluate the disk footprint of \textsc{LiveDB} and contrast it with the storage overhead observed in the Geth-based implementation used in Sonic. Since original database does not prune automatically, \textsc{LiveDB} consumes only a fraction of the disk space, without incurring any additional cost. 

\item[\textbf{\em Claim-II: Storage Efficiency of Hash Organization.}] We evaluate the disk footprint of \textsc{ArchiveDB} and contrast it with the storage overhead observed in the Geth-based implementation used in Sonic. Since original database is not pruned, its content is equivalent to archive, while our new design radically reduces disk space consumption. 

\end{description}

\subsection{Experimental Setup} \paragraph{Hardware Configuration}

The experiments were conducted on a bare-metal infrastructure provided by a server provider. The machine offered 32 virtual CPUs and 128 GB of RAM. Local NVMe storage was used to ensure optimal I/O performance. The software stack ran on Debian Linux, using Go version 1.23.

\paragraph{Dataset and Replay Methodology}

We utilized the tool from \cite{Kim2021off-the-chain}, which operates in two distinct phases. In the initial phase, the tool synchronizes with a live blockchain and captures the necessary input data to enable deterministic re-execution of transactions. This phase was applied to the Sonic mainnet, which has been active since 2025. Importantly, the recording phase was excluded from performance measurements.

In the second phase, the tool replays the recorded blockchain data at maximum achievable speed, constrained only by hardware and software limits. This replay mechanism allows us to extract key performance indicators, such as transactions per second and aggregate gas usage over time.

\paragraph{Baseline Comparison}

We conducted three primary experimental runs using the Sonic client. The first run employed the original database layer derived from Ethereum’s Geth client, which utilizes LevelDB for key-value storage. In the second and third run, we replaced this layer with our redesigned database implementation for \textsc{LiveDB} and \textsc{ArchiveDB}. This setup enabled a direct comparison of throughput and storage characteristics between the legacy and proposed architectures.

\subsection{Throughput Evaluation}

To assess the performance implications of removed read amplification, we carried out a controlled experiment where we first replayed  previously captured blockchain data with original database forked from Ethereum (geth), then we repeated the experiment with \textsc{LiveDB}. Throughout the replay process, we recorded the block height to compute the corresponding transactions per second (tx/s). 

\begin{figure*}
    \centering
    \includegraphics[width=1\textwidth]{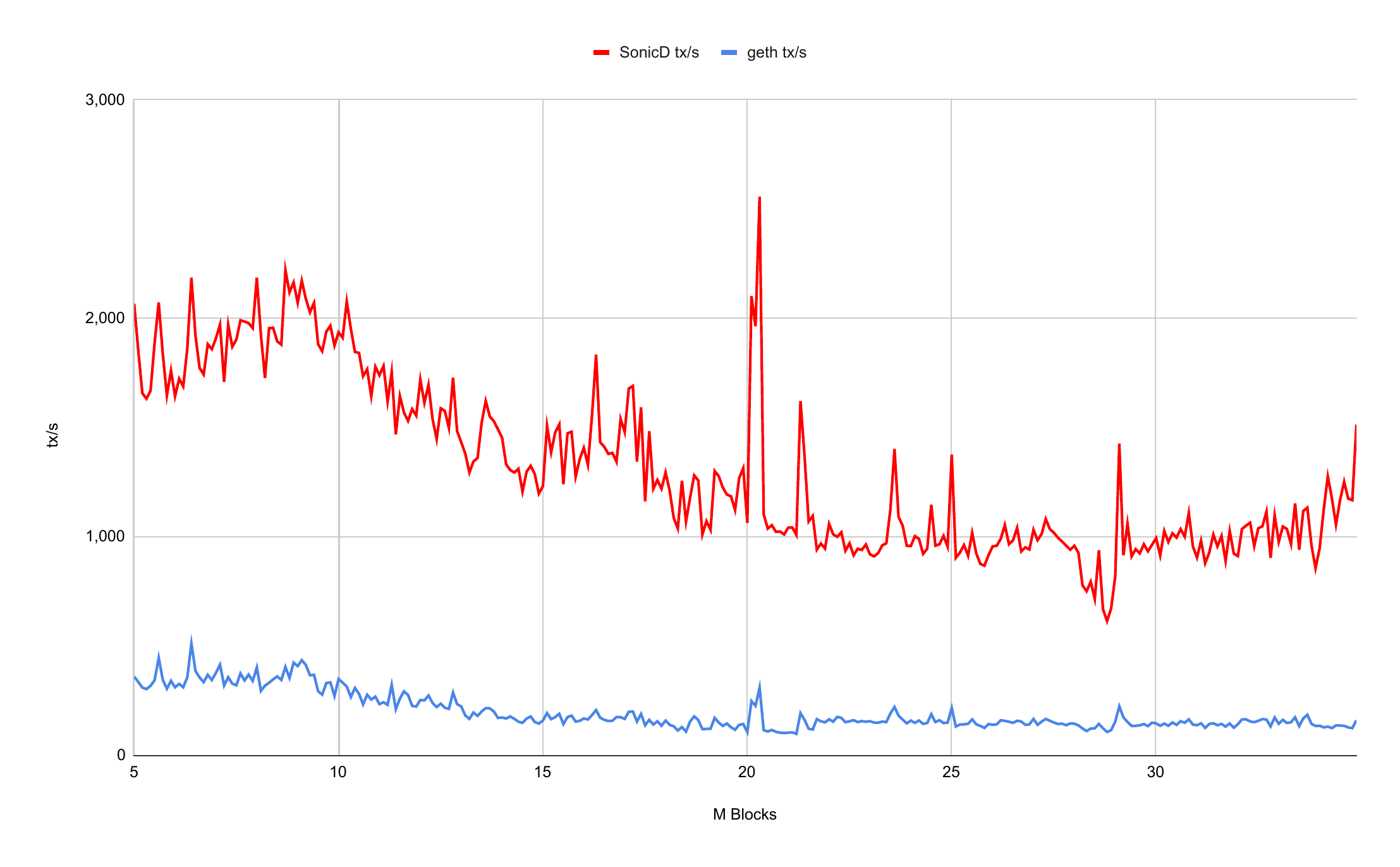}
    \caption{Transaction throughput comparison between original database (geth) and our new system (SonicDB). The new system is about 10 times faster, featuring ~1.479 tx/s at block 17M vs. 165 tx/s for the MPT. }
    \label{fig:throughput}
\end{figure*}

Figure~\ref{fig:throughput} presents a comparative analysis of transaction throughput between the original database (geth) and \textsc{LiveDB} (SonicDB). At the midpoint of the experiment, corresponding to block height 17 million, the baseline (geth) configuration sustained a throughput of approximately 
165 tx/s. In contrast, SonicDB achieved a throughput of approximately 1,479tx/s at the same block height, indicating almost a tenfold improvement in execution performance. 

To put the numbers above to another perspective, synchronizing the whole blockchain using the SonicDB took about 22hours, while using Geth, it took above six days.

It is important to note that results for the initial segment of the experiment, spanning blocks 0 to 5 million, are omitted from the chart. During this early phase, SonicDB exhibited throughput approaching 17,000tx/s. However, this elevated performance is likely attributable to the fact that the entire dataset fit within memory, thereby bypassing disk I/O operations. As such, these measurements do not reflect typical database-bound performance and are excluded to maintain the integrity of the comparison.

Both systems exhibited a modest reduction in performance at growing block heights. This degradation is primarily attributed to the growing database footprint, which adversely affects cache locality and increases the frequency of disk I/O operations. Nevertheless, transaction throughput remained largely stable across the duration of the experiment.  

Our system design supports dynamic activation and switching of the \textsc{ArchiveDB} component, which introduces a potential overhead due to the need to propagate data simultaneously to both \textsc{LiveDB} and \textsc{ArchiveDB}. To evaluate the effect of this dual-write mechanism on transaction throughput, we replicated our earlier experiment with the archive enabled. Through continuous monitoring, we observed that transaction throughput remained stable, with no measurable degradation.

\begin{figure*}
    \centering
    \includegraphics[width=1\textwidth]{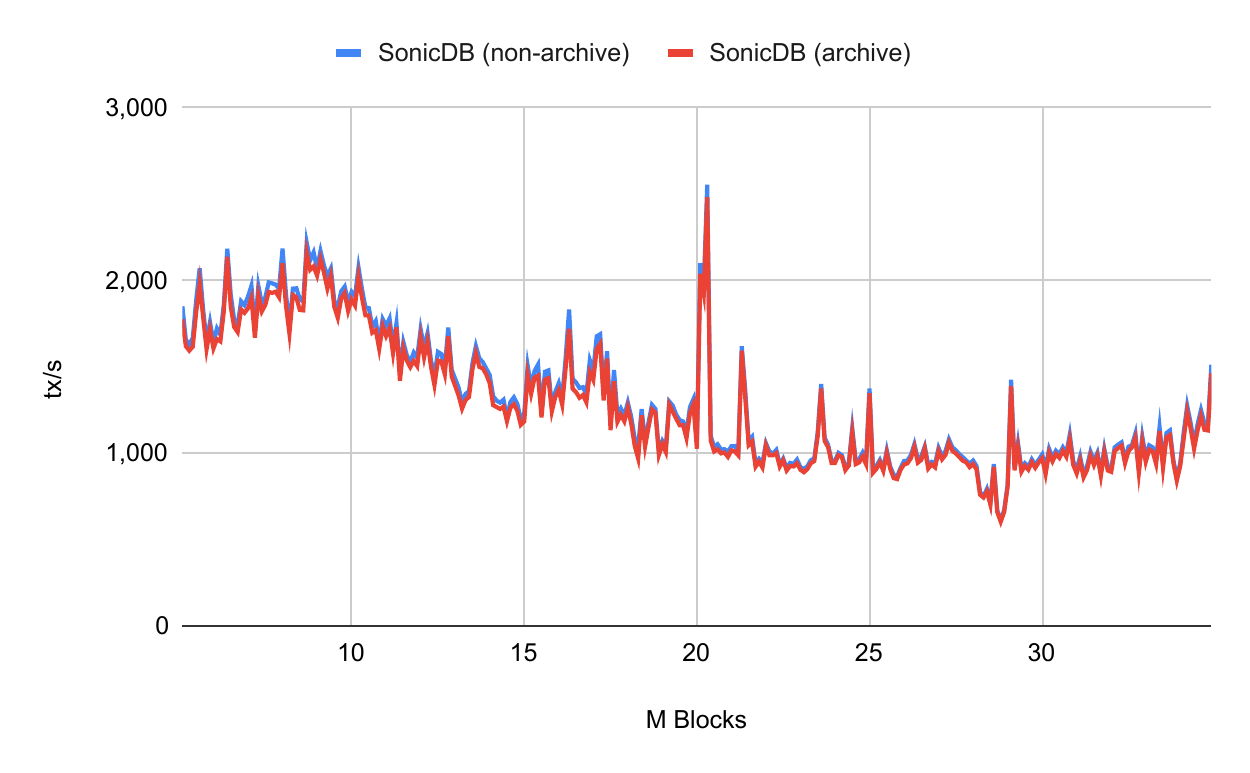}
    \caption{Throughput comparison of LiveDB and Archive; the Archive does not incur any significant slowdown.}
    \label{fig:carmen-archive}
\end{figure*}

Figure~\ref{fig:carmen-archive} presents a comparative plot of throughput across \textsc{LiveDB} and the archive configuration. The near-complete overlap of performance curves indicates that the addition of archival storage has negligible impact on execution efficiency. The observed throughput progression closely mirrors that of Figure~\ref{fig:throughput}, where only \textsc{LiveDB} was active. These findings suggest that enabling archival functionality does not compromise system performance.

\subsection{Storage Overhead Evaluation}

\begin{figure*}
    \centering
    \includegraphics[width=1\textwidth]{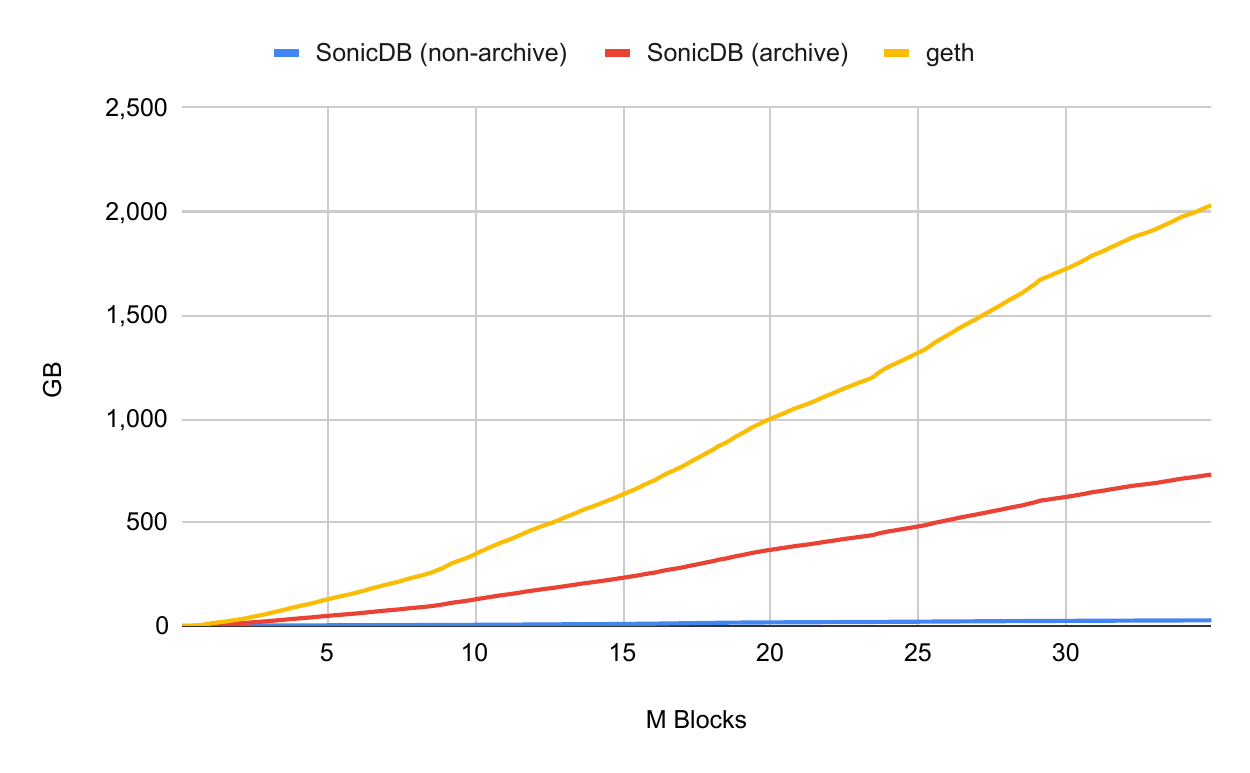}
    \caption{Disk space allocation comparison between original database (geth) and our new system (SonicDB) running both as archive and non-archive. The new system needed 27GB for LiveDB and 730GB for Archive, while geth needed above 2TB, at block 34M.}
    \label{fig:disk}
\end{figure*}

While running the experiments from the previous section for three configurations: geth, \textsc{LiveDB} and \textsc{ArchiveDB} we in addition measured the disk space consumption. This time it shows how much space can be saved when the cost-free automatic pruning is available with \textsc{LiveDB}, and how the hash organization improvement reduces the disk space for \textsc{ArchiveDB}. A summary of the comparative results is provided in Table~\ref{fig:disk}.

The evaluation revealed a substantial disparity in storage requirements between the original geth, and our proposed SonicDB architecture. At block height 34 million, the Geth-based MPT consumed above 2TB of disk space, whereas SonicDB's \textsc{LiveDB} required only 27GB, representing a reduction of nearly 99\% in storage overhead, and the \textsc{ArchiveDB} needed 730GB, i.e. about third od Getht's requirements. Furthermore, the data suggests that the gap between SonicDB and Geth is increasing throughout block heigh, i.e. the scale of the problem will grow with enlarging the blockchain. 

\section{Related Work}
\label{sec:related}

Read and write amplification in trie-based data structures has been a central topic in blockchain systems research. 
Raju et al.~\cite{raju2018mlsm} introduced Merkelized Log-Structured Merge Trees (mLSM), which combine Merkle trees with LSM architectures to reduce write overhead.
Ponnapalli et al.~\cite{ponnapalli2019scalable} explored distributed Merkle Patricia Tries (MPTs) across multiple nodes to enhance scalability. Additional efforts have focused on optimizing Merkle tree variants for space and performance~\cite{dahlberg2016efficient, haider2018compact, ostersjo2016sparse}. Despite these innovations, our empirical findings indicate that trie traversal remains heavily constrained by underlying database latency and read amplification, limiting practical performance gains. 
Layered Merkle Patricia Tries (LMPTs) \cite{choi2023lmpt}, split the state into three layers: delta, intermediate, and snapshot tries. This design enables non-blocking I/O and reduces read/write amplification by caching recent updates in memory. However, LMPTs require multiple root hashes and are not compatible with Ethereum’s single-root model. 
LVMT \cite{li2023lvmt} employs Authenticated Multipoint Evaluation Trees (AMTs) to reduce commitment update costs. While LVMT supports constant-time root updates and efficient sharding, its complexity and reliance on large precomputed elliptic curve tables limit practical deployment. 
SALT \cite{yang2025salt}, flattens sparse subtries using history-independent hash tables to improve space efficiency. It minimizes I/O by storing commitments in memory and propagating updates through a vector-packed trie. However, its multi-structure design complicates implementation and limits archival support. 
NOMT \cite{habermeier2024nomt} leverages SSD parallelism by batching data into 4KB pages and storing trie nodes in rootless B-trees. This approach significantly reduces I/O operations and supports high-throughput execution, but introduces archive overhead and depends on parallel EVM execution.
NURGLE \cite{he2024nurgle} introduces a novel denial-of-service attack that targets specifically MPT and its unbalance nature that may inflate resource consumption. Through systematic evaluation, the authors demonstrate how intermediate node proliferation degrades performance.
QMDB \cite{zhang2025qmdb} keeps Merkleization entirely in memory using fixed-size subtrees called “twigs.” It achieves constant-time read/write amplification and efficient batched writes, though it suffers from low historical lookup efficiency and high metadata overhead.
Finally, Ethereum Binary Trees, formalized in EIP-7864 \cite{eip7864}, revisit binary trie structures with simplified node types and improved proof efficiency. While promising in terms of zero-knowledge friendliness and quantum resistance, their increased read amplification and pending hash function standardization delay adoption.

Cache-Tries, proposed by Prokopec~\cite{hashtrie}, offer a concurrent hash trie design that achieves constant-time operations through a quiescently consistent caching mechanism. This structure incorporates single-linked lists that maintain pointers to trie nodes at each level, ordered from leaf to root. The cache facilitates rapid lookups and insertions by enabling bottom-up traversal, with fallback to slower top-down search when necessary.
Privacy-preserving storage has also been explored in blockchain contexts. BlockStorage~\cite{blockstorage} introduces a model tailored for IoT environments, utilizing federated extreme learning machines to classify and store frequently accessed blocks while safeguarding user privacy. This approach improves query efficiency and ensures secure handling of sensitive data.
Feng et al.~\cite{Feng23} propose a compact storage scheme for blockchain transaction data that supports secure verification. Their method employs advanced data structures to minimize storage requirements while maintaining integrity and auditability, offering a promising direction for scalable blockchain storage.

Concurrency has been extensively studied as a means to improve smart contract execution~\cite{dickerson2017adding, zhang2018enabling, anjana2019efficient, shrey2019dipetrans, sarrar2019transaction, saraph2019empirical}. These approaches typically involve constructing execution graphs with forks and joins, replicated across network nodes. While this technique removes database access from the critical execution path, it does not eliminate latency associated with state retrieval.

Efforts to accelerate key-value storage systems have yielded notable results. Faster~\cite{chandramouli2018faster} spans both disk and memory using a log-structured design. Song et al.~\cite{song2018multipath} introduced bulk retrieval methods atop RocksDB to optimize SSD performance. Papagiannis et al.~\cite{papagiannis2018efficient} reported 4–6× speedups using memory-mapped key-value stores. Ouaknine et al.~\cite{ouaknine2017optimization} demonstrated significant gains through fine-tuned RocksDB configurations. Wu et al.~\cite{wu2015lsm} proposed the LSM-trie, achieving an order-of-magnitude improvement with LevelDB and RocksDB. Lepers et al.~\cite{lepers2019kvell} designed KVell, a database optimized for SSDs, while Agarwal et al.~\cite{agarwal2015succinct} explored querying compressed data to reduce disk usage without sacrificing performance.

The Ethereum ecosystem explored sharding as a strategy to partition state and distribute computational load~\cite{ethereum2020sharding}. Sharding divides accounts into clusters, allowing independent processing across shards. Recent developments have shifted scalability efforts to Layer 2 (L2) chains, with Layer 1 (L1) serving as a slower settlement layer. Although, recent advances in L1 chains made these sharding ideas obsolete. 

State management optimizations have also emerged. The Geth client introduced state snapshots via a separate index\footnote{\url{https://github.com/ethereum/go-ethereum/pull/20152}}, enabling faster access to recent state. Live pruning of the MPT, supported by the Klaytn Foundation\footnote{\url{https://docs.klaytn.foundation/docs/build/tools/indexers/}}, selectively removes obsolete nodes to reduce storage overhead. The Path-Based scheme\footnote{\url{https://ethresear.ch/t/ethereum-path-based-storage-model-and-newly-inline-state-prune/14932}}, implemented in Geth\footnote{\url{https://github.com/ethereum/go-ethereum/issues/25390}}, introduces a mutable MPT variant with embedded pruning capabilities.
Ethereum also employs a Snapshot mechanism\footnote{\url{https://github.com/ethereum/go-ethereum/pull/20152}} that retains the last 128 blocks in a flattened key-value format alongside the MPT. While this reduces reliance on the trie for recent history, it introduces architectural complexity and additional storage requirements.

Verkle Tries~\cite{kuszmaul2018verkle}, combining vector commitments with trie structures, were proposed as a successor to MPTs in Ethereum. They offer dramatically smaller proof sizes and improved zero-knowledge compatibility~\cite{oberst2025stateless}. Despite their theoretical advantages, Verkle Tries were ultimately deprecated due to performance concerns and lack of quantum resistance.

\section{Conclusion}
\label{sec:conclusion}
We have presented a novel Ethereum compatible stateDB data structure starting fast forkless blockchains. Our stateDB is based on a mutable tree structure that separates expensive hash maintenance from the stateDB operations. Our stateDB exhibits a 
10x increase in throughput and a 99\% reduction in storage, or 3x reduction when archival data is kept. 

\bibliographystyle{plain}
\bibliography{main}

\end{document}